\documentclass[referee]{raa}            

\usepackage{graphicx,times}             
\usepackage{natbib}
\usepackage{amssymb,amsmath}
\usepackage{pdflscape}
\usepackage{color}
\bibpunct{(}{)}{;}{a}{}{,}

\usepackage[a4paper=true,pagebackref=true]{hyperref}
\hypersetup{colorlinks = true, linkcolor = green, anchorcolor = red, citecolor = blue, filecolor = red, pagecolor = red, urlcolor = red}

\begin{document}

   \title{An Explanation for 13 consecutive days activities of Mrk 421}

 \volnopage{ {\bf 2012} Vol.\ {\bf X} No. {\bf XX}, 000--000}
   \setcounter{page}{1}

\author{Y.G. Zheng\inst{1,2,3,4}, C.Y. Yang\inst{1,4}, Shi-Ju Kang\inst{5} and J.M. Bai\inst{1,4}}
\institute{Yunnan Observatories, Chinese Academy of Sciences, Kunming 650011, China; {\it ynzyg@ynu.edu.cn~~baijinming@ynao.ac.cn}\\
           \and
           Department of Physics, Yunnan Normal University, Kunming, 650092, China\\
           \and
           Shandong Provincial Key Laboratory of Optical Astronomy and Solar-Terrestrial Environment-Shandong University, Weihai, 264209, China\\
           \and
           Key Laboratory for the Structure and Evolution of Celestial Objects, Chinese Academy of Sciences\\
           \and
           Department of Physics and Electronics Science, Liupanshui Normal University, Liupanshui, Guizhou, 553004, China\\
\vs \no
   {\small Received XXXX; accepted XXXX}
}

\abstract{
It is surprising to find a fact for migration in the peak positions of synchrotron spectra energy distribution component during in the activity epochs of Mrk 421, accompanying with an orphan flaring at the X-ray and GeV-TeV $\gamma$-ray bands. A geometric interpretation and standard shock or stochastic acceleration models of blazar emission have difficulty reproducing these observed behaviours. The present paper introduces a linear acceleration by integrating the reconnection electric field into the particle transport model for the observed behaviours of Mrk 421. We note that the strong evidence for evolution of multi-wavelength spectral energy distribution characteristic by shifting the peak frequency, accompanying with an orphan flaring at the X-ray and GeV-TeV $\gamma$-ray bands provides an important electrostatic acceleration diagnostic in blazar jet. Assuming suitable model parameters, we apply the results of the simulation to the 13-day flaring event in 2010 March of Mrk 421, concentrating on the evolution of multi-wavelength spectral energy distribution characteristic by shifting the peak frequency. It is clear that the ratio of the electric field and magnetic field strength plays an important role in temporal evolution of the peak frequency of synchrotron spectral energy distribution component. We suggest the electrostatic acceleration responsible for the evolution of multi-wavelength spectral energy distribution characteristic by shifting the peak frequency is reasonable. Based on the model results, we issue that the peak frequency of the synchrotron spectral energy distribution component may denote a temporary characteristic of blazars, rather than a permanent one.
}
   \authorrunning{Zheng et al.}            
   \titlerunning{An Explanation for 13 consecutive days activities of Mrk 421}  

   \maketitle

\section{Introduction}\label{sec:intro}

Blazars are the radio-loud active galactic nuclei (AGN). In the most accepted scenario, their continuum emissions extending from radio to $\gamma$-ray bands arise from the relativistic jet. This jet emerges from supermassive black holes and beams the emission at the observer's line of sight (e.g., \citealt{1986ApJ...310..317G, 1995PASP..107..803U}). Hence, the observed spectra are subjected to the effects of Doppler boosting \citep{1979ApJ...232...34B}. In general, multi-wavelength observations show the spectral energy distribution (SED) of blazars in the $\log\nu-\log\nu F_{\nu}$ space exhibits a double-peak shape \citep{1998MNRAS.299..433F}. It is believed that the low-energy peak of the SED is attributed to synchrotron emission from extreme relativistic electrons and/or positrons in the jet \citep{1998AdSpR..21...89U}. The origin of the high-energy component remains an open issue. The lepton model suggests that it may be attributed to the inverse Compton (IC) scattering of the extremely relativistic electrons (e.g., \citealt{1974ApJ...188..353J, 1989ApJ...340..181G, 1992A&A...256L..27D, 2016MNRAS.457.3535Z,2017ApJS..228....1Z}).

Most of blazars appear as luminous sources characterized by noticeable and rapid flux variability at all observed frequencies. Generally, prominent X-ray and $\gamma$-ray flares trend to be correlated with low frequency flares (e.g., \citealt{2017Natur.552..374R}). Due to the observational limitations, it is difficult to obtain a detailed picture of the time-varying emission spectrum. Previous efforts to study temporary spectrum from a blazar have concentrated mainly on the production of variability via either time-dependent particle distribution (e.g., \citealt{1996ApJ...461..657B, 1999MNRAS.306..551C, 2002ApJ...581..127B, 2006A&A...453...47K, 2011ApJ...728..105Z, 2015A&A...573A...7W, 2016ApJ...824..108L}) or some physical (e.g., \citealt{1979ApJ...232...34B, 1985ApJ...298..114M, 1991ApJ...377..403C, 1997A&A...320...19M, 1998A&A...333..452K, 2007A&A...462...29G, 2008A&A...491L..37M}) and/or geometrical parameters changing (e.g., \citealt{1992A&A...255...59C, 1992A&A...259..109G}). In recent years, the unprecedented study of multi-wavelength observation campaigns obtain the spectra of blazars as close in time as possible (e.g., \citealt{2015A&A...578A..22A, 2016ApJ...819..156B, 2016ApJ...827...55K, 2017ApJ...834....2A, 2018A&A...620A.181A}). While models endeavor to describe the temporary characteristics of the emission spectrum \citep{2015A&A...578A..22A, 2016ApJ...819..156B}, migration in the peak positions of the synchrotron emission SED component during activity epochs remains an open issue.

Both the diffusive shock acceleration (e.g. \citealt{1978MNRAS.182..147B, 1983SSRv...36...57D, 2014ApJ...780...87M, 2018MNRAS.478.3855Z}) and stochastic acceleration (e.g., \citealt{1985A&A...143..431S, 1989ApJ...336..243S, 2004ApJ...610..550P, 2011ApJ...728..105Z, 2015ApJ...808L..18A, 2017MNRAS.464.4875B}) are potentially efficient mechanisms for producing energetic particles from a plasma flow with strong shocks. In spite of that the mechanisms, including the injection, acceleration and cooling of particles \citep{2002A&A...386..833G}, with the possible intervention of shock waves \citep{1985ApJ...298..114M, 2001ApJ...554....1S} or turbulence \citep{2014ApJ...780...87M}, are still being debated on producing the unpredictable variability, previous model efforts can simulate the properties of these flares (e.g., \citealt{2011ApJ...728..105Z, 2012ApJ...760...69N, 2014A&A...571A..83P}). There is a class of $\gamma$-ray flares that occurs with little correlated variability that is detected at longer wavelengths (e.g., \citealt{2004ApJ...601..151K, 2013A&A...552A..11R}). Since the theoretical photon spectral reproducing in standard shock or stochastic acceleration models are calculated in the synchrotron self-Compton (SSC) model frame, the mentioned models are difficult to explain this pure orphan flare. For the purpose of application proposed model, in the context we expand the orphan flare to include the flares at both X-ray and $\gamma$-ray bands.

It is generally believed that the magnetohydrodynamic (MHD) instabilities of a Poynting-flux dominated plasma flow results in formation of current sheets where the magnetic reconnection is triggered \citep{1993ApJ...419..111E, 1998ApJ...493..291B, 2006A&A...450..887G}. Particle acceleration in reconnection has been used to interpret the prompt phenomena of $\gamma$-ray bursts (e.g., \citealt{2013ApJ...769...69B, 2014ApJ...782...92Z, 2016MNRAS.459.3635B, 2017MNRAS.468.3202B}), in addition to the high-energy non-thermal emission from pulsar wind nebulae (e.g., \citealt{2003ApJ...591..366K, 2012ApJ...754L..33C, 2013ApJ...770..147C, 2014ApJ...785L..33P, 2017A&A...607A.134C}) and AGN jets (e.g., \citealt{1992A&A...262...26R, 2009MNRAS.395L..29G, 2010MNRAS.402.1649G, 2011MNRAS.413..333N, 2012MNRAS.420..604N, 2013MNRAS.431..355G, 2016MNRAS.462.3325P, 2019MNRAS.482...65C}). While the particle-in-cell (PIC) simulations \citep{2014ApJ...783L..21S, 2014PhRvL.113o5005G, 2017ApJ...843L..27W} reveal several difficulties in understanding how the reconnection produces efficient particle acceleration, one can suggest that linear acceleration by the reconnection electric field is the simplest and most fundamental acceleration mechanism \citep{2018MNRAS.476.3902K}. In principle, the magnetic reconnection offers natural locations, within the diffusion region where the magnetic field is small and reverse, in which the electric field can exceed the magnetic field \citep{2013ApJ...770..147C}. These results indicate the electrostatic acceleration can drive up the particle energy to a level significantly overrun the possible values when only the shock and stochastic acceleration are considered. In these scenarios, we anticipate extreme incident behavior in the quasi-spherical emission region of the blazar jet to occur.

In the present context, we attempt to account for the temporary characteristic of the emission spectrum by introducing an electrostatic acceleration mechanism. The aim of paper is to find the distinctive energy spectrum characteristics for the electrostatic acceleration in the blazar jet. The present paper is organized as follows. In Section 2, we take the rate of the linear electrostatic acceleration into account in the model by assuming a constant electric field strength. In Section 3, we discuss the model parameters and establish a ratio of the electric field and the magnetic field strengths reflecting the electrostatic acceleration rate. In Section 4, we estimate the maximum Lorentz factor of the particles in the model. Based on the expected maximum Lorentz factor, we deduce the maximum energy of the synchrotron peak, and we propose an important electrostatic acceleration diagnostic in the blazar jet. In Section 5, we apply the model to the 13-day flaring event in 2010 March of Mrk 421. This application concentrates on the electrostatic acceleration responsible for the evolution of the multi-wavelength SED characteristic by shifting the peak frequency; a discussion is given in Section 6.

Throughout the paper, all of the parameters are calculated in the co-moving frame. And then, we transfer them from the co-moving frame to the observed frame by taking relativistic beaming effect into account. We assume the Hubble constant $H_{0}=75$ km s$^{-1}$ Mpc$^{-1}$, the dimensionless numbers for the energy density of matter $\Omega_{\rm M}=0.27$, the dimensionless numbers of radiation energy density $\Omega_{\rm r}=0$, and the dimensionless cosmological constant $\Omega_{\Lambda}=0.73$.

\section{The proposed model}
The photon spectra in the current context are calculated by the model within the lepton model frame through both synchrotron emission and the IC scattering. In the proposed model, we basically follow the approach of \cite{2019ApJ...873....7Z} to calculate the electron spectrum, and we introduce electrostatic acceleration to the first-order gain rate. In our approaches, since the magnetic reconnections occurs in a natural location, we do not take the actual electrostatic acceleration process. Phenomenologically, we treat the electrostatic acceleration as a mechanism of gain the energy in the transport equation, that is, the electrons pick up the energy form the electric fields.

Assuming an electric field of strength $E$ is generated in the magnetic reconnection region around the shock, we estimate the electrostatic acceleration rate experienced by the electrons with the magnitude of the electron charge, $e$, due to electric field crossing in the comoving frame as follows
\begin{equation}
\dot{p}_{\rm elec}=eE\,.
\label{Eq:1}
\end{equation}
Combining the momentum gain rate with shock acceleration, $\dot{p}_{\rm sh}$, we can establish the first-order momentum gain rate
\begin{equation}
\dot{p}_{\rm gain}=\dot{p}_{\rm elec}+\dot{p}_{\rm sh}=A_{0}m_{e}c\,,
\label{Eq:2}
\end{equation}
where, $A_{0}$ is the first-order momentum gain rate constant in the unit of $\rm s^{-1}$ with $A_{0}=A_{\rm sh}+A_{\rm elec}$, $m_{e}$ the electron mass, and $c$ the speed of light. In this scenario, we rebuild the basic transport equation as follows
\begin{eqnarray}
\frac{\partial f(p,t)}{\partial t}&=&\frac{1}{p^{2}}\frac{\partial}{\partial p}\biggl\{p^{2}\biggl[D_{0}m_{e}cp\frac{\partial f(p,t)}{\partial p}-A_{0}m_{e}c f(p,t)\nonumber\\&+&\frac{B_{0}p^{2}}{m_{e}c}f(p,t)\biggr]\biggr\}-\biggl(\frac{C_{0}m_{e}c}{p}+\frac{F_{0}p}{m_{e}c}\biggr)f(p,t)\nonumber\\&+&\frac{\dot{N}_{0}\delta(p-p_{0})}{4\pi p_{0}^{2}}\,,
\label{Eq:3}
\end{eqnarray}
where, $f(p, t)$ is the isotropic, homogeneous phase-space density, $p$ is the particle momentum, $D_{0}$ is the momentum diffusion rate constant, $B_{0}$ is the momentum-loss rate constant, $C_{0}$ is the shock-regulated escape rate constant, $F_{0}$ is the Bohm diffusive escape rate constant, $\dot{N}_{0}$ is the continual injection rate in the units of $~p^{-1}\rm cm^{-3}~s^{-1}$, $p_{0}$ is the characteristic injection momentum, and $\delta(p)$ is the Dirac's distribution function.

Using a Green's function to solve the stationary particle transport equation (see e.g., \citealt{2016ApJ...833..157K, 2019ApJ...873....7Z}), we can obtain the closed-form solution for the electron Green's function, $N_{G}(\gamma, \gamma_{0})$, showing the number density distribution of electrons in the Lorentz factor, $\gamma$, space with the Gamma function, $\Gamma(x)$, as follows
\begin{eqnarray}
N_{G}(\gamma, \gamma_{0})&=&\frac{\dot{N}_{0}m_{e}c\Gamma(\mu-\sigma+0.5)}{\hat{B}D_{0}\Gamma(1+2\mu)\gamma_{0}^{2}}\biggl(\frac{\gamma}{\gamma_{0}}\biggr)^{\frac{\hat{A}}{2}} e^{-\frac{\hat{B}(\gamma^{2}-\gamma_{0}^{2})}{4}}\nonumber\\&\times&M_{\sigma, \mu}(\frac{\hat{B}\gamma_{\rm 1}^{2}}{2})W_{\sigma, \mu}(\frac{\hat{B}\gamma_{\rm 2}^{2}}{2})\,,
\label{Eq:4}
\end{eqnarray}
where, we let $\hat{A}=A_{0}/D_{0}$, $\hat{B}=B_{0}/D_{0}$, $\hat{C}=C_{0}/D_{0}$, and $\hat{F}=F_{0}/D_{0}$. The model assumes that the injected particles have a mono-energetic distribution, and the parameter $\gamma_{0}$ is a characteristic Lorentz factor of the injected particles. The Whittaker functions $M_{\sigma, \mu}(\frac{\hat{B}\gamma_{\rm 1}^{2}}{2})$ and $W_{\sigma, \mu}(\frac{\hat{B}\gamma_{\rm 2}^{2}}{2})$ are defined by the parameter $\sigma=1+\hat{A}/4-\hat{F}/(2\hat{B})$, and $\mu=0.25[(2+\hat{A})^{2}+4\hat{C}]^{1/2}$ with $\gamma_{1}=\rm min[\gamma, \gamma_{0}]$ and $\gamma_{2}=\rm max[\gamma, \gamma_{0}]$.

We can use the number density distribution of electrons determined by Eq. (\ref{Eq:4}) to calculate the emission intensity of the synchrotron, $I_{\rm syn}(\nu)$, and the IC scattering, $I_{\rm ic}(\nu)$ in the SSC model frame (e.g., \citealt{2001A&A...367..809K, 2019ApJ...873....7Z}). Taking the absorption effect of the extragalactic background light (EBL) into account, we can calculate the flux density at Earth (e.g., \citealt{2011ApJ...728..105Z})
\begin{equation}
F_{\rm obs.}(\nu_{\rm obs.})=\frac{\pi\delta^{3}(1+z)r_{s}^{2}}{d_{L}^{2}}\biggl[I_{\rm syn}(\nu)+I_{\rm ic}(\nu)\biggr]\times e^{-\tau(\nu,z)}\;,
\label{Eq:5}
\end{equation}
where, $\delta$ is the Doppler factor (e.g., \citealt{1979rpa..book.....R}), $z$ the redshift of the source, $r_{s}$ the size of the blob, $d_{L}$ the luminosity distance, and $\tau(\nu,z)$ the absorption optical depth (e.g., \citealt{2005ApJ...618..657D}). The relationship between the frequency, $\nu_{\rm obs.}$, at observer's frame and the frequency, $\nu$, at co-moving frame is given by $\nu_{\rm obs.}=\delta\nu/(1+z)$.
\section{The model parameters}
The application of the model requires the specification of both the particle spectra parameters (i.e., $\dot{N_{0}}$, $D_{0}$, $\gamma_{0}$, $\hat{A}$, $\hat{B}$, $\hat{C}$, and $\hat{F}$), and the jet parameters ($B$, $\delta$, and $r_{s}$). Since we introduce the electrostatic acceleration to the first-order gain rate, we should reestablish the parameter relationships, rather than relying on the model parameters determined in our early work. Based on the equations presented by \cite{2019ApJ...873....7Z} and the Eqs. (\ref{Eq:1}) and (\ref{Eq:2}) in the current context, we find
\begin{equation}
\hat{A}=\hat{A}_{\rm sh}+\hat{A}_{\rm elec}=\frac{9\eta\xi}{4\sigma_{\rm mag}}+\frac{E}{B}\frac{3\eta}{\sigma_{\rm mag}}\;,
\label{Eq:6}
\end{equation}
\begin{equation}
\hat{B}=5.54\times10^{-13}\frac{\eta u}{\sigma_{\rm mag}}(\frac{B}{0.1~\rm G})^{-1}\;,
\label{Eq:7}
\end{equation}
\begin{equation}
\hat{C}=\frac{3\eta}{\omega\sigma_{\rm mag}}\;,
\label{Eq:8}
\end{equation}
and
\begin{equation}
\hat{F}=8.74\times10^{-26}\eta^{2}\sigma_{\rm mag}^{-1}(\frac{r_{s}}{10^{17}~\rm cm})^{-2}(\frac{B}{0.1~\rm G})^{-2}\;,
\label{Eq:9}
\end{equation}
where, $\eta$ is a dimensionless parameter, $\xi$ is the efficiency factor of shock acceleration, $\sigma_{\rm mag}$ is the magnetization parameter, $B$ is the local magnetic field strength, $u$ is the summation of both the magnetic field and the soft photon field energy density, and $\omega$ is a dimensionless constant of order unity.

In our proposed approach, we require increasing the summation of both the magnetic field and the soft photon field energy density as free particle spectral parameters; this is based on the free parameters of \cite{2019ApJ...873....7Z}. Once the values of particle spectral parameters are established, we can derive the momentum diffusion rate constant as follows,
\begin{equation}
D_{0}=\frac{4\sigma_{T}u}{3m_{e}c\hat{B}}\;,
\label{Eq:10}
\end{equation}
where, $\sigma_{T}$ is the with Thomson cross section, the magnetization parameter, $\sigma_{\rm mag}$, with
\begin{equation}
\sigma_{\rm mag}=\frac{3\eta D_{0}m_{e}c}{eB}\;,
\label{Eq:11}
\end{equation}
and the dimensionless timescale constant, $\omega$, is as follows
\begin{equation}
\omega=\frac{3\eta}{\hat{C}\sigma_{\rm mag}}\;.
\label{Eq:12}
\end{equation}
Next, we can obtain the averaged soft photon field, $u_{\rm ph}$, with
\begin{equation}
u_{\rm ph}=u-\frac{B^{2}}{8\pi}\;,
\label{Eq:13}
\end{equation}
and the ratio of the electric field and magnetic field strength, $E/B$, as follows
\begin{equation}
\frac{E}{B}=\frac{\hat{A}\sigma_{\rm mag}}{3\eta}-\frac{3}{4}\xi\;,
\label{Eq:14}
\end{equation}
\section{Radiation energy of maximum particles}
The transport equation in the current context takes the shock, electrostatic, stochastic acceleration, synchrotron and IC scattering loss, as well as particle injection and escape into account. A dynamic equilibrium is generated by a kind of competition among the acceleration, cooling, injection and escape of particles from the shock region. Neglecting the particle injection and escape, we can estimate the maximum Lorentz factor of the particles, $\gamma_{\rm max}$, by examining the Fokker-Plank ``drift'' coefficient, $\langle dp/dt\rangle=D_{0}m_{e}c(3+\hat{A}-\hat{B}\gamma_{\rm max}^{2})=0$, which yields
\begin{equation}
\gamma_{\rm max}^{2}=\gamma_{\rm e, stoch}^{2}+\gamma_{\rm e, sh}^{2}+\gamma_{\rm e, elec}^{2}\;,
\label{Eq:15}
\end{equation}
It is convenient to obtain the theoretical equilibrium momentum for combination of stochastic acceleration and momentum loss by balancing the stochastic momentum gain rate, $\dot{p}_{\rm stoch}$, with the momentum-loss rate, $\dot{p}_{\rm loss}$, so that the equation
\begin{equation}
(\dot{p}_{\rm stoch}+\dot{p}_{\rm loss})|_{p=p_{\rm e, stoch}}=0\;,
\label{Eq:16}
\end{equation}
can be established. The equilibrium Lorentz factor for stochastic acceleration versus momentum loss is therefore given by
\begin{equation}
\gamma_{\rm e, stoch}=\frac{p_{\rm e, stoch}}{m_{e}c}=\sqrt{\frac{9D_{0}m_{e}c}{4\sigma_{T}u}}\;.
\label{Eq:17}
\end{equation}
With the same approach, we can establish the following equations
\begin{equation}
(\dot{p}_{\rm sh}+\dot{p}_{\rm loss})|_{p=p_{\rm e, sh}}=0\;,
\label{Eq:18}
\end{equation}
and
\begin{equation}
(\dot{p}_{\rm elec}+\dot{p}_{\rm loss})|_{p=p_{\rm e, elec}}=0\;,
\label{Eq:19}
\end{equation}
which yields the equilibrium Lorentz factor for shock acceleration versus momentum loss,
\begin{equation}
\gamma_{\rm e, sh}=\frac{p_{\rm e, sh}}{m_{e}c}=\sqrt{\frac{9\xi eB}{16\sigma_{T}u}}\;,
\label{Eq:20}
\end{equation}
and the equilibrium Lorentz factor for electrostatic acceleration versus momentum loss,
\begin{equation}
\gamma_{\rm e, elec}=\frac{p_{\rm e, elec}}{m_{e}c}=\sqrt{\frac{3eE}{4\sigma_{T}u}}\;,
\label{Eq:21}
\end{equation}
respectively.

The Larmor timescale is the minimum timescale for acceleration of the particles via energetic collisions with MHD waves,
\begin{equation}
t_{\rm L}(p)=\frac{r_{\rm L}}{c}=\frac{p}{eB}\;,
\label{Eq:22}
\end{equation}
where $r_{\rm L}$ is the Larmor radius. Equating the Larmor timescale with the timescale of momentum loss
\begin{equation}
[t_{\rm L}(p)+t_{\rm loss}(p)]|_{p=p_{\rm MHD}}=0\;,
\label{Eq:23}
\end{equation}
yields a critical Lorentz factor,
\begin{equation}
\gamma_{\rm MHD}=\frac{p_{\rm MHD}}{m_{e}c}=\sqrt{\frac{3eB}{4\sigma_{T}u}}\;,
\label{Eq:24}
\end{equation}

Since both shock and stochastic acceleration are mediated by interactions between the electrons and the MHD wave, we expect neither the $\gamma_{\rm e, stoch}$ nor the $\gamma_{\rm e, sh}$, to exceed the critical Lorentz factor, $\gamma_{\rm MHD}$. These assumptions give
\begin{equation}
D_{\rm 0, max}=\frac{eB}{3m_{e}c}\;,
\label{Eq:25}
\end{equation}
and $\xi_{\rm max}=4/3$, respectively. On the other hand, we deduce the combination of these two processes cannot accelerate electron with a Lorentz factor that exceeds the critical Lorentz factor, that is
\begin{equation}
(\dot{p}_{\rm stoch}+\dot{p}_{\rm sh}+\dot{p}_{\rm loss})|_{p=p_{\rm MHD}}\leq0\;.
\label{Eq:26}
\end{equation}
This relation yields
\begin{equation}
(\frac{D_{0}}{D_{\rm 0, max}}+\frac{3}{4}\xi)\leq1\;.
\label{Eq:27}
\end{equation}
Substituting Eqs. (\ref{Eq:17}), (\ref{Eq:20}), (\ref{Eq:21}) and (\ref{Eq:25}) into Eq. (\ref{Eq:15}) yields
\begin{eqnarray}
\gamma_{\rm max}&=&\sqrt{\frac{3eB}{4\sigma_{T}u}(\frac{D_{0}}{D_{\rm 0, max}}+\frac{3}{4}\xi+\frac{E}{B})}\nonumber\\
&=&7.36\times10^{7}\biggl(\frac{B}{0.1~\rm G}\biggr)^{1/2}\biggl(\frac{u}{0.01~\rm erg~cm^{-3}}\biggr)^{-1/2}\nonumber\\&\times&\biggl(\frac{D_{0}}{D_{\rm 0, max}}+\frac{3}{4}\xi+\frac{E}{B}\biggr)^{1/2}\;,
\label{Eq:28}
\end{eqnarray}

The characteristic energy of the synchrotron emission peak is (e.g., \citealt{1979rpa..book.....R})
\begin{equation}
E_{\rm syn, peak}(\gamma)=\gamma^{2}m_{e}c^{2}\frac{B}{B_{\rm crit}}=0.12~\rm MeV \biggl(\frac{\gamma}{10^{7}}\biggr)^{2}\biggl(\frac{B}{0.1~\rm G}\biggr)\;,
\label{Eq:29}
\end{equation}
where, the $B_{\rm crit}$ is critical magnetic field with $B_{\rm crit}=4.41\times10^{13}~\rm G$. We can substitute Eq. (\ref{Eq:28}) into Eq. (\ref{Eq:29}) to obtain the photon energy of the synchrotron peak, given by
\begin{eqnarray}
E_{\rm syn, peak}(\gamma_{\rm max})&=&\gamma_{\rm max}^{2}m_{e}c^{2}\frac{B}{B_{\rm crit}}\nonumber\\
&=&6.50~\rm MeV\biggl(\frac{B}{0.1~\rm G}\biggr)^{2}\biggl(\frac{u}{0.01~\rm erg~cm^{-3}}\biggr)^{-1}\nonumber\\&\times&\biggl(\frac{D_{0}}{D_{\rm 0, max}}+\frac{3}{4}\xi+\frac{E}{B}\biggr)\;.
\label{Eq:30}
\end{eqnarray}
Substituting Eq. (\ref{Eq:27}) into Eq. (\ref{Eq:30}) yields
\begin{eqnarray}
E_{\rm syn, peak}^{\rm max}(\gamma_{\rm max})&\leq&6.50~\rm MeV\biggl(\frac{B}{0.1~\rm G}\biggr)^{2}\biggl(\frac{u}{0.01~\rm erg~cm^{-3}}\biggr)^{-1}\nonumber\\&\times&\biggl(1+\frac{E}{B}\biggr)\;.
\label{Eq:31}
\end{eqnarray}
Eq. (\ref{Eq:31}) indicates that the maximum photon energy of the synchrotron peak, $E_{\rm syn, peak}^{\rm max}(\gamma_{\rm max})$ depends on the magnetic field strength, $B$, radiation field energy density, $u_{\rm ph}$, and electric field strength, $E$.

\begin{figure}
	\centering
		\includegraphics[angle=0,width=10cm]{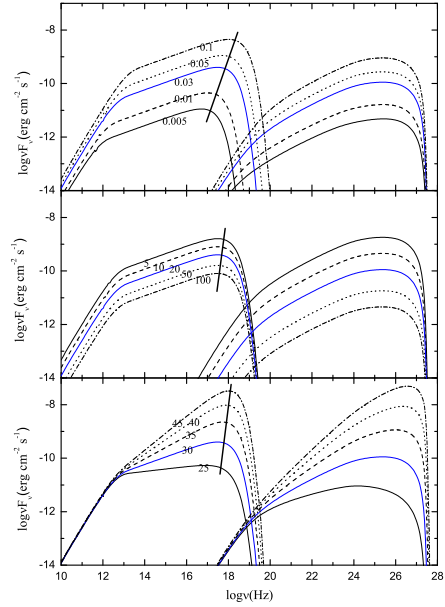}	
	\caption{The changes in the multi-wavelength SED by varying the physical parameters. The top panel shows the magnetic field strength $B$, the middle panel shows the summation of both the magnetic field and the soft photon field energy density $u$, which indicates the varying the radiation field energy density, $u_{\rm ph}$, when the parameter $B$ is fixed. The bottom panel shows the dimensionless parameter $\hat{A}$, which indicates the varying ratio of the electric field and magnetic field strength, $E/B$, when the parameters $B$, and $u$ are fixed. The labels near the curves represent the values of the parameter. The thick lines are plotted by the peak frequency of the synchrotron SED component. The blue curve denotes the baseline of the multi-wavelength SED with the chosen of parameters $\gamma_{0}=1000$, $\dot{N}_{0}=2.0\times10^{19}~\rm p^{-1}~cm^{-3}~s^{-1}$, $\xi=0.1$, $\eta=1$, $u=20~\rm erg~cm^{-3}$, $\hat{A}=30$, $\hat{B}=8.54\times10^{-11}$, $\hat{C}=84$, $\delta=32$, $B=0.03~\rm G$, and $r_{s}=2.1\times10^{16}~\rm cm$.}
	\label{Fig:1}
\end{figure}

It is important to note that we can employ the three most common approaches to explain the evolution of the multi-wavelength SED characteristic by shifting the peak frequency. The three approaches are (1) varying the magnetic field strength; (2) varying the radiation field energy density; and (3) varying the electric field strength. As can be seen in Eqs (\ref{Eq:10}) and (\ref{Eq:14}), regardless of changing either the magnetic field strength or the radiation field energy density, these changes result in a substantial influence on the value of the electrostatic field in the proposed model of in the current context. The changes in the multi-wavelength SED as a result of varying the physical parameters $B$, $u$ and $\hat{A}$ are plotted in Figure {\ref{Fig:1}}. One can see that the migration in the peak positions of the synchrotron SED component dominates in the evolution of the multi-wavelength SED. Phenomenally, we anticipate that some incidental behaviours can be expected from the emission spectra. Neglecting the influence on the other physical conditions, if we focus on varying either the magnetic field strength or the radiation field energy density, since the dynamic equilibrium of the system reestablishes, we can find a continuous variability covering energies from high to low. Alternatively, if we focus on varying the electric field strength, the efficiency acceleration contains the electron populations around on the equilibrium energy, we can expect the shapes of these SED bumps to change. In addition, the flux variability is remarkable and the change of spectral slopes is significant at the X-ray and GeV-TeV $\gamma$-ray bands; however, as a result of that the low energy electrons are dominated by the electrostatic acceleration process, the variability is minor or not significant at the low frequency end of synchrotron component. In these scenarios, we issue that there is strong evidence for the evolution of the multi-wavelength SED characteristic by shifting the peak frequency, and this is accompanied with by an orphan flaring at the X-ray and GeV-TeV $\gamma$-ray bands. Consequently, these observations provide an important electrostatic acceleration diagnostic in blazar jets.

\section{Application to Mrk 421}
Mrk 421 is a nearby active galaxy at a redshift of $z=0.031$ \citep{1991rc3..book.....D}, with a pair of relativistic jets flowing in opposite directions that is closely aligned to our line of sight. 
It is interesting to find the significant evolution of the multi-wavelength SED characteristic by shifting the peak frequency has been exhibited during activity epochs (e.g., \citealt{2010ApJ...724.1509U, 2015A&A...578A..22A, 2016ApJ...819..156B}). During these epochs, the archival multi-wavelength observations of Mrk 421 show the synchrotron and the IC scattering peak in the quiescent state located at $\sim1~\rm keV$ and $\sim100~\rm GeV$, respectively (e.g., \citealt{2011ApJ...736..131A}). On the other hand, the 13 consecutive days (from MJD 55265 to MJD 55277) activity of Mrk 421 occurring in March 2010 show an orphan flaring at the X-ray and GeV-TeV $\gamma$-ray bands \citep{2015A&A...578A..22A}.

In order to attempt to understand the nature of the migration in the peak positions of the SED, in the current context, we apply the proposed model to the 13-day flaring event in March 2010 of Mrk 421, concentrating on the electrostatic acceleration responsible for the evolution of multi-wavelength SED characteristic by shifting the peak frequency. We first establish the values of the model parameters. Our approach for reproducing the multi-wavelength spectrum from Mrk 421 sets $\eta=1$, $\xi=0.1$, and $\delta=32$ in all of the observation epochs. Since the adopted multi-wavelength SEDs are averaged daily, we argue that our approximation is valid as long as the variability timescale, $t_{\rm var}$, is less than 1-day. In this scenario, we constrain the size of the emission region by the relation $r_{s}\sim c\delta t_{\rm var}/(1+z)=2.1\times10^{16}~\rm cm$ with $t_{\rm var}=2.25\times10^{4}~\rm s$ in all of the calculation epochs. The other model parameters $\dot{N_{0}}$, $\gamma_{0}$, $\hat{A}$, $\hat{B}$, $\hat{C}$, $u$, and $B$ are varied until a reasonable qualitative fit to the multi-wavelength spectra data is obtained.

\begin{figure}
\centering
		\includegraphics[angle=0,width=17cm]{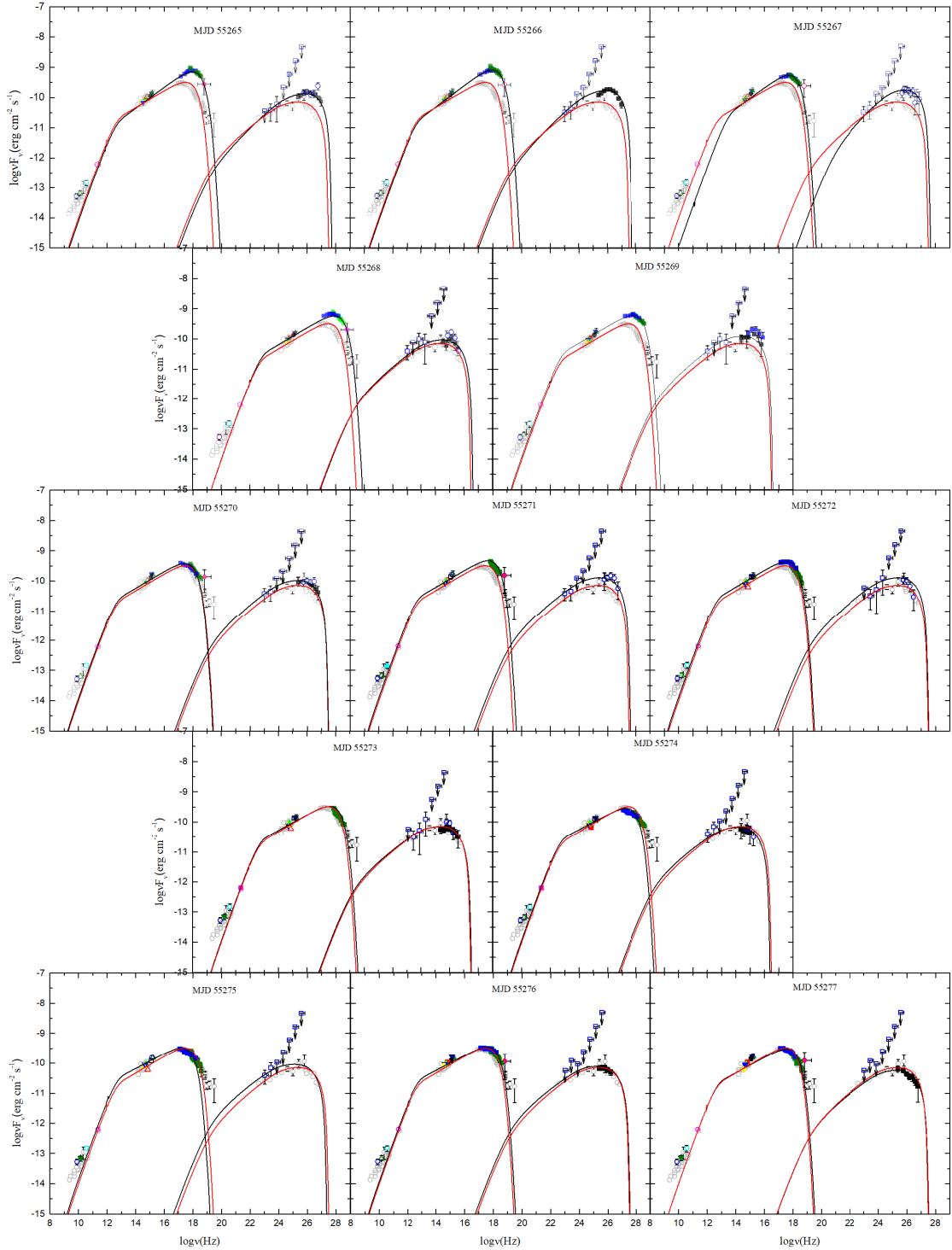}
\caption{Comparisons of model spectra with observed data for the BL Lac object Mrk 421. The color symbols show the daily SEDs during 13 consecutive days (from MJD 55262 to MJD 55277) of activity taken from \cite{2015A&A...578A..22A}. The \emph{Fermi} upper limits are show as blue open squares with down arrow. The gray circles in the background of each panel show the averaged SED from the 2009 MW campaign reported in \cite{2011ApJ...736..131A}, which is a good representation of the nonflaring SED of Mrk 421. The model-derived spectra that fit the daily SEDs are plotted with red curves in each panel, while spectra that fit the nonflaring SED are plotted with black curves.}
  \label{Fig:2}
\end{figure}

In Figure {\ref{Fig:2}}, we plot the model spectrum along with the simultaneous multi-wavelength SED observations reported by \cite{2015A&A...578A..22A} in the epoch of 2010 March from Mrk 421. The plots include a comparison of the nonflaring (or typical) SED from the 2009 MW campaign observed in \cite{2011ApJ...736..131A}. The corresponding physical parameters of the model spectrum for the 13-day flaring event in 2010 March and the nonflaring state of Mrk 421 are reported in Table {\ref{Table:1}}. We can estimate the specific amount of the shock and the electric field by using $\hat{A}_{\rm sh}=9\eta\xi/(4\sigma_{\rm mag})$, and $\hat{A}_{\rm elec}=3E\eta/(B\sigma_{\rm mag})$. The results are also listed in Table {\ref{Table:1}}. We can see that the electrostatic acceleration dominates in the 13 consecutive days activities. The inferred values of electric field in the reconnection layer are also included in Table 1. It can be seen that, for the strong flares at the beginning of the activity epochs, the condition $E\gtrsim B$ is in accord with rapid magnetic reconnection, which can result in efficient electrostatic acceleration. The exceptions are the tear of flare activity. For these weak flares, we obtain $E<B$. Since the intensities of these particular flare activities barely exceeded the level of the quiescent emission, we issue that the results are reasonable in the sense that strong electrostatic acceleration is not required to interpret the spectrum observed during that 13 consecutive days activities.

\begin{table*}
\centering
\tiny
\caption{The proposed model parameters\label{Table:1}.}
\tabcolsep 1.20mm
\begin{tabular}{cccc ccccc ccccc ccccc ccc}
\hline \hline
Date & $\dot{N_{0}}$ & $\gamma_{0}$ & $\xi$  & $\eta$ & $\hat{A}$ & $\hat{B}$ & $\hat{C}$ &  $u$ & $B$ & $\delta$ & $r_{s}$ & $\hat{F}$ & $D_{0}$ & $\sigma_{\rm mag}$ & $\omega$  &  $E$ & $\hat{A}_{\rm sh}$& $\hat{A}_{\rm elec}$\\
MJD & $[\rm p^{-1}~cm^{-3}~s^{-1}]$ & ... & ... & ... &	... & $[10^{-11}]$ &	... & $[\rm erg~cm^{-3}]$ & $[\rm G]$ & ... & $[10^{16}~\rm cm]$ & $[10^{-22}]$ & $[\rm s^{-1}]$ & ... & ... & $[\rm G]$ & ... & ...\\
\hline

typical         & $2.0\times10^{19}$ 	&	1000	&	0.1	&	1.0 	&	30 	&	8.54	 &	84.0&	25.38	&	 0.030	&	32 	&	 2.1 	&	4.01	&	 $9.65\times10^{3}$ 	&	 0.055 	&	0.65 	&	0.014 	&	4.103 & 25.92	\\
55265	        & $3.5\times10^{19}$ 	&	1000	&	0.1	&	1.0 	&	20 	&	2.54	 &	56.0&	25.38	&	 0.030	&	32 	&	 2.1 	&	1.19	&	 $3.24\times10^{4}$ 	&	 0.185 	&	0.29 	&	0.035	&	1.222 & 18.82	\\
55266	        & $3.5\times10^{19}$ 	&	1000	&	0.1	&	1.0 	&	19 	&	2.54	 &	53.2&	25.38	&	 0.030	&	32 	&	 2.1 	&	1.19	&	 $3.24\times10^{4}$ 	&	 0.185 	&	0.31 	&	0.033	&	1.222 & 17.82	\\
55267	        & $3.0\times10^{18}$ 	&	3000	&	0.1	&	1.0 	&	20 	&	2.54	 &	56.0&	25.38	&	 0.023	&	32 	&	 2.1 	&	2.15	&	 $2.35\times10^{4}$ 	&	 0.175 	&	0.31 	&	0.025	&	1.292 & 18.76	\\
55268	        & $3.2\times10^{19}$ 	&	1000	&	0.1	&	1.0 	&	30 	&	4.50	 &	84.0&	25.38	&	 0.040	&	32 	&	 2.1 	&	1.59	&	 $1.83\times10^{4}$ 	&	 0.078 	&	0.46 	&	0.028	&	2.885 & 27.16	\\
55269	        & $3.2\times10^{19}$ 	&	1000	&	0.1	&	1.0 	&	30 	&	5.50	 &	84.0&	25.38	&	 0.038	&	32 	&	 2.1 	&	2.04	&	 $1.50\times10^{4}$ 	&	 0.067 	&	0.53 	&	0.023	&	3.347 & 26.67	\\
55270	        & $2.2\times10^{19}$ 	&	1000	&	0.1	&	1.0 	&	30 	&	9.50	 &	84.0&	25.38	&	 0.030	&	32 	&	 2.1 	&	4.46	&	 $8.68\times10^{3}$ 	&	 0.049 	&	0.72 	&	0.012	&	4.562 & 25.44	\\
55271	        & $4.0\times10^{19}$ 	&	1000	&	0.1	&	1.0 	&	30 	&	5.50	 &	84.0&	25.38	&	 0.030	&	32 	&	 2.1 	&	2.58	&	 $1.50\times10^{4}$ 	&	 0.085 	&	0.42 	&	0.023	&	2.642 & 27.37	\\
55272	        & $3.0\times10^{19}$ 	&	1000	&	0.1	&	1.0 	&	30 	&	7.50	 &	84.0&	25.38	&	 0.030	&	32 	&	 2.1 	&	3.52	&	 $1.10\times10^{4}$ 	&	 0.063 	&	0.57 	&	0.016	&	3.603 & 26.42	\\
55273	        & $2.8\times10^{19}$ 	&	1000	&	0.1	&	1.0 	&	35 	&	7.50	 &	98.0&	25.38	&	 0.032	&	32 	&	 2.1 	&	3.30	&	 $1.10\times10^{4}$ 	&	 0.059 	&	0.52 	&	0.019	&	3.843 & 31.18	\\
55274	        & $1.6\times10^{19}$ 	&	1000	&	0.1	&	1.0 	&	37 	&	15.0	 &103.6&	25.38	&	 0.032	&	32 	&	 2.1 	&	6.61	&	 $5.49\times10^{3}$ 	&	 0.029 	&	0.98 	&	0.009	&	7.694 & 29.37	\\
55275	        & $1.6\times10^{19}$ 	&	1000	&	0.1	&	1.0 	&	37 	&	18.0	 &103.6&	25.38	&	 0.032	&	32 	&	 2.1 	&	7.93	&	 $4.58\times10^{3}$ 	&	 0.024 	&	1.19 	&	0.007	&	9.223 & 27.80	\\
55276	        & $3.1\times10^{19}$ 	&	1000	&	0.1	&	1.0 	&	37 	&	8.0	     &103.6&	25.38	&	 0.032	&	32 	&	 2.1 	&	3.52	&	 $1.03\times10^{4}$ 	&	 0.055 	&	0.53 	&	0.019	&	4.101 & 32.94	\\
55277	        & $2.4\times10^{19}$ 	&	1000	&	0.1	&	1.0 	&	35	&	8.0	     & 98.0&	25.38	&	 0.032	&	32 	&	 2.1 	&	3.52	&	 $1.03\times10^{4}$ 	&	 0.055 	&	0.56 	&	0.018	&	4.101 & 30.94	\\

\hline
\end{tabular}
\end{table*}

\begin{figure}
\centering
\includegraphics[width=15.0 cm]{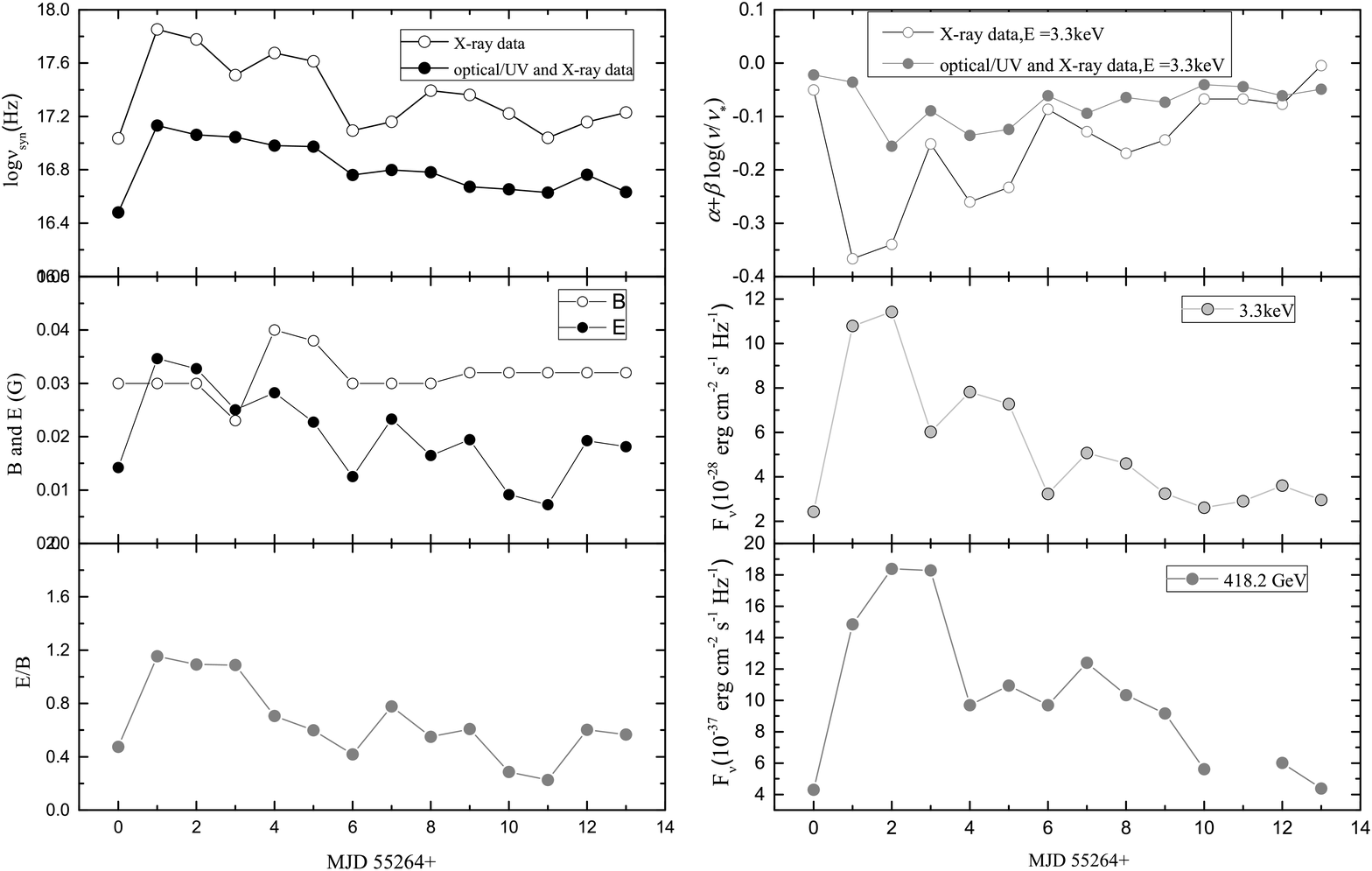}
\caption{Left panels: Temporal evolution of the peak frequency of the synchrotron SED component (top panel), both the electric field and the magnetic field strength (middle panel), and the ratio of the electric field and the magnetic field strength (bottom panel) during the 13-day flaring event in 2010 March of Mrk 421; Right panels: Temporal evolution of the spectral slopes of synchrotron emission (top panel), the X-ray (middle panel) and $\gamma$-ray (bottom panel) fluxes during the 13-day flaring event in 2010 March of Mrk 421. The peak location of the synchrotron SED component is estimated by fitting a log-parabolic model in Eq. ({\ref{Eq:31}}). The associated ratio of the electric field and the magnetic field strength is computed using the model parameters. The plot assumes the typical state at the epoch of MJD 55264. We note that the $\gamma$-ray fluxes are absent from observation in the day of MJD 55275.}
  \label{Fig:3}
\end{figure}

We now take the evolution of the multi-wavelength SED characteristic by shifting the peak frequency into account. Since very high-energy (VHE) photons, general $E_{\gamma}>0.1$ TeV, from the source are attenuated by photons from the EBL (e.g., \citealt{1992ApJ...390L..49S, 2002A&A...386....1K, 2004A&A...413..807K, 2005ApJ...618..657D, 2006ApJ...648..774S, 2008A&A...487..837F, 2009ApJ...697..483R, 2010ApJ...712..238F}), we deduce the peak position of the high-energy SED component from Mrk 421 depends on the $\gamma\gamma$ attenuation effect. In this scenario, we do not consider the peak position of the IC scattering SED component; instead, we estimate the peak location of the synchrotron SED component using a reasonable smooth interpolation or extrapolation model (e.g., \citealt{2004A&A...413..489M, 2005ApJ...630..130B, 2009A&A...501..879T, 2010ApJ...724.1509U}). We introduce a log-parabolic model with
\begin{equation}
F(\nu)=k\biggl(\frac{\nu}{\nu_{*}}\biggr)^{-\alpha-\beta\log(\nu/\nu_{*})}\;,
\label{Eq:32}
\end{equation}
in the $\nu-F(\nu)$ space (e.g., \citealt{2016ApJ...819..156B}). we utilize the model to fit to the X-ray data alone, and to both the UV/optical and X-ray data. In this model, $\alpha$ and $\beta$ are free parameters, while $\nu_{*}$ is a fixed parameter at which $\alpha$ equals the local power-law photon index. The $\beta$ parameter exhibits the deviation of the spectral slope away from $\nu_{*}$. The evolution of the peak frequency of the synchrotron SED component in the observed epoch is reported in Table {\ref{Table:2}}. The corresponding peak frequencies of the synchrotron SED component for each activity are plotted in Figure {\ref{Fig:3}}. The plot includes the comparison of both the electric field and the magnetic field strength, the ratio of the electric field and the magnetic field strength, the spectral slopes of synchrotron emission, and the light curve of X-ray and $\gamma$-ray. We note that the agreement between the migration in the peak positions of the synchrotron SED component and evolution of the ratio of the electric field and magnetic field strength is excellent during the activity epochs. Since the magnetic field strength in the observation epoch experiences minor changes, and the summation of both the magnetic field and the soft photon field energy density is fixed, we suggest that it is reasonable to conclude that the electrostatic acceleration is responsible for the evolution of the multi-wavelength SED characteristic by shifting the peak frequency. We are interesting to compare the ratio of the electric field and the magnetic field strength with the spectral slopes of synchrotron emission, and the light curve of X-ray and $\gamma$-ray. One can find that when the ratio increases, the spectral slopes of synchrotron emission enlarge, and the X-ray and $\gamma$-ray fluxes strengthen. These are the consequence of that the electrostatic acceleration in an electric field tends to harden the particle distribution, which enhances the high-energy component of resulting synchrotron emission spectrum.

\begin{table*}[htbp]
\caption{Results of the log-parabolic model$^{a}$ to fit the X-ray data alone, and both the UV/optical and X-ray data\label{Table:2}.}
\centering
\scriptsize
\begin{tabular}{lccccccccccccc}
\hline\hline
    &  \multicolumn{6}{c}{X-ray data} & & \multicolumn{6}{c}{optical/UV and X-ray data}\\
\cline{2-7} \cline{9-14} \\
 date & $k$          & $\alpha$  & $\beta$ & $\nu_{*}$ &$\chi^{2}$ &$\nu_{syn, p}$ & & $k$          & $\alpha$  & $\beta$ & $\nu_{*}$ &$\chi^{2}$ &$\nu_{syn, p}$ \\
 MJD  & [$10^{-28}$] & ...       & ...     & [Hz]      &...        &[$10^{17}$~Hz] & & [$10^{-28}$] & ...       & ...     & [Hz]      &...        &[$10^{16}$~Hz] \\
\hline\\
typical & 1.259 &	-1.575	&	0.274 &	$2.173\times10^{12}$ & 0.007  &	1.086&      &	1.000	&	-1.533	&	0.172 &	$1.309\times10^{9}$	& 0.008	&3.022\\
55265	& 1.000	&	-1.785	&	0.527 &	$1.622\times10^{15}$ & 0.002  &	7.118&  	&	7.079	&	-1.412	&	0.221 &	$4.721\times10^{11}$& 0.024 &13.52\\
55266   & 1.000	&	-1.802	&	0.490 &	$8.279\times10^{14}$ & 0.002  &	5.987&  	&	1.000	&	-1.692	&	0.227 &	$1.358\times10^{11}$& 0.015 &11.55\\
55267	& 0.832	&	-1.494	&	0.244 &	$2.512\times10^{12}$ & 0.001  &	3.241&  	&	1.000	&	-1.447	&	0.157 &	$1.791\times10^{9}$	& 0.002 &11.12\\
55268	& 1.000	&	-1.708	&	0.414 &	$2.541\times10^{14}$ & 0.001  &	4.737&  	&	1.000	&	-1.672	&	0.218 &	$7.129\times10^{10}$& 0.009 &9.582\\
55269	& 1.000	&	-1.688	&	0.385 &	$1.327\times10^{14}$ & 0.001  &	4.111&  	&	1.000	&	-1.636	&	0.209 &	$4.645\times10^{10}$& 0.007 &9.402\\
55270	& 1.000	&	-1.556	&	0.237 &	$5.035\times10^{11}$ & 0.004  &	1.243&  	&	1.000	&	-1.512	&	0.171 &	$2.606\times10^{9}$	& 0.003 &5.768\\
55271	& 1.000	&	-1.678	&	0.284 &	$2.786\times10^{12}$ & 0.003  &	1.445&  	&	1.000	&	-1.660	&	0.214 &	$3.837\times10^{10}$& 0.003 &6.294\\
55272	& 1.000	&	-1.778	&	0.434 &	$1.563\times10^{14}$ & 0.002  &	2.480&  	&	1.778	&	-1.745	&	0.275 &	$6.166\times10^{11}$& 0.010 &6.038\\
55273	& 1.000	&	-1.774	&	0.450 &	$1.905\times10^{14}$ & 0.002  &	2.303&  	&	1.000	&	-1.757	&	0.248 &	$1.297\times10^{11}$& 0.007 &4.693\\
55274	& 1.000	&	-1.499	&	0.268 &	$3.631\times10^{12}$ & 0.001  &	1.668&  	&	1.000	&	-1.495	&	0.166 &	$1.377\times10^{9}$	& 0.003 &4.509\\
55275	& 1.000	&	-1.602	&	0.271 &	$1.730\times10^{12}$ & 0.003  &	1.093&  	&	1.000	&	-1.539	&	0.177 &	$2.858\times10^{9}$	& 0.005 &4.249\\
55276	& 1.000	&	-1.481	&	0.220 &	$3.311\times10^{11}$ & 0.003  &	1.441&  	&	0.794	&	-1.455	&	0.146 &	$2.265\times10^{8}$	& 0.003 &5.788\\
55277	& 2.667	&	-1.411	&	0.299 &	$1.578\times10^{13}$ & 0.008  &	1.697&  	&	1.000	&	-1.560	&	0.182 &	$3.972\times10^{9}$	& 0.006 &4.285\\
\hline\\
\end{tabular}
\flushleft{$^{a}$NOTE-The fitting is obtained by considering the photon energy $\nu$ to be the independent variable. Transformation the log-parabolic model into $\log\nu-\log\nu F_{\nu}$ space, we can obtain a relation $\log\nu F_{\nu}=c_{0}+c_{1}\log\nu+c_{2}(\log\nu)^{2}$, where $c_{0}=-\beta(\log\nu_{*})^{2}+\alpha\log\nu_{*}+\log k$, $c_{1}=2\beta\log\nu_{*}-\alpha+1$, and $c_{2}=-\beta$. The parameters $c_{1}$ and $c_{2}$ can localize the synchrotron SED peak by $\nu_{\rm syn, p}=-c_{1}/(2c_{2})$.}
\end{table*}

\section{Conclusion and discussion}
It is believed that determining the jet microphysics from the SED of blazar is a challenging problem of inversion. In a phenomenological view, it is interesting and important to find the exclusive characteristic of the emission energy spectrum for tracing the jet microphysics. The present paper introduces a linear acceleration by integrating the reconnection electric field into the particle transport model which has been proposed by \cite{2019ApJ...873....7Z}. We expected to find the observed evidence for the electrostatic acceleration in the blazar jet. In the model, the ratio of the electric field and the magnetic field strength reflecting the electrostatic acceleration rate is established. On the basis of the model's expected maximum Lorentz factor, we deduce the maximum photon energy of the synchrotron peak in the SSC model frame. The result indicates that the energy of the synchrotron peak depends on the ratio of the electric field and the magnetic field strength. In order to justify the migration in the peak positions of the synchrotron SED component, we test the theoretical SED effects of changes in the parameters. The tests provide strong evidence for the evolution of multi-wavelength SED characteristic by shifting the peak frequency, with an accompanying orphan flaring at the X-ray and GeV-TeV $\gamma$-ray bands. These results provide an important electrostatic acceleration diagnostic in blazar jets. Assuming suitable model parameters, we apply the results of the simulation to the 13-day flaring event in March 2010 of Mrk 421, concentrating on the evolution of the multi-wavelength SED characteristic by shifting the peak frequency. It is clear that the ratio of the electric field and the magnetic field strength play an important role in the temporal evolution of the peak frequency of the synchrotron SED component. Therefore, we consider it reasonable to suggest the electrostatic acceleration is responsible for the evolution of multi-wavelength SED characteristic by shifting the peak frequency.

Phenomenally, the model effort of \cite{2015A&A...578A..22A} for the 13 consecutive days activities of Mrk 421 focus on the correlated variability at the X-ray and $\gamma$-ray bands without variability at the optical/UV bands. Their model paradigm confirms that the one-zone SSC model can describe the SED of each day reasonably well. Alternatively, the flaring activity can be reproduced by a two-zone SSC model, where one zone is responsible for the quiescent emission, and the other zone contributes to the daily variable emission occurring at X-ray and $\gamma$-ray bands \citep{2015A&A...578A..22A}. In the context model, we concentrate on the evolution of the multi-wavelength SED characteristic by shifting the peak frequency. To render the nonlinear effect of SSC process, we assume a constant soft photon fields to instead of the synchrotron emission field. With the adopted parameters $B = 0.03~\rm G$, it results in $u_{B}\sim3.6\times10^{-5}~\rm erg~cm^{-3}$. Based on the model results, we can also estimate the synchrotron emission field $u_{syn}\sim10^{-4}~\rm erg~cm^{-3}$. If we directly include the synchrotron emission field into the transport equation, we find $u=u_{B}+u_{syn}\sim10^{-4}~\rm erg~cm^{-3}$. This value is five orders of magnitude less than the constant soft photon fields of the model assumed with $u=25.38 ~\rm erg~cm^{-3}$. It is believed that a lower soft photon field results to larger equilibrium Lorentz factor. The calculated synchrotron emission field induces the equilibrium Lorentz factor $\gamma_{e}$ around $10^{7}\sim10^{8}$. This is far from the equilibrium Lorentz factor that are required by the observed synchrotron peak energy. In order to obtain a corresponding synchrotron peak energy, we assume a large value for soft photon fields to establish the particle distributions. It is important to assume a plane shock front propagating along a cylindrical jet. A particle energy distribution is shaped by experiencing a kind of acceleration, cooling, injection and escape processes around the shock front, and subsequently drift away into the downstream flow, in which they emit most of their energy. Since the context introduces a shock-induced magnetic reconnection scenario, we must emphasize the aim of model assumed soft photon fields is to shape a particle energy distribution around the shock front, whereas, the emission spectra are strictly calculated in the frame of SSC process.

In the context, we do not provide detailed insights on what is the dominate acceleration mechanism in the reconnection region. However, it is interesting to find that all of the island contraction \citep{2006Natur.443..553D}, anti-reconnection between colliding islands \citep{2010ApJ...714..915O, 2014ApJ...783L..21S}, and curvature drift acceleration \citep{2014PhRvL.113o5005G, 2015ApJ...806..167G} are able to transfer the available magnetic energy into a population of non-thermal ultra-relativistic electrons via the induced large-scale electric field. Actually, the high energy particles are regulated and focused towards the reconnection layer midplane (e.g., \citealt{2007A&A...472..219C}). Once the particles are immersed the layer, they suffer from weak radiative losses, but strong coherent electric fields \citep{2011ApJ...737L..40U}. On the contrary, the ambient magnetic field exceeds the electric field in the surrounding region. In this scenario, the layer can be treated as a linear accelerator \citep{2013ApJ...770..147C}. Since magnetic reconnections offer natural regions where the electric field is stronger than the magnetic field, we adopt a simple linear acceleration mechanism by the reconnection electric field. On the basis of assuming a constant electric field strength in the region, we propose a possible explanation for the origin of the temporal evolution on the peak frequency of the synchrotron emission SED component.

It is convenient to compare the relative contributions from shock acceleration and electrostatic acceleration in the region of the dissipated region. The model presented in this work suggests the ratio of the electric field and magnetic field strength in a limited range from 0.2 to 1.2 during in the activity epochs. Since we adopt the shock acceleration efficiency factor $\xi=0.1$ in the context, we can conclude that the first-order momentum gain by electrostatic acceleration is a dominated mechanism at the dissipation region in the jet. However, we should emphasized that shock none the less plays an important role in regulating the particles escape, and therefore it is an essential composition in the proposed model, despite the momentum gain by shock acceleration is negligible. Actually, in the environment where the magnetic reconnection is efficient, so as to provide the electric field, the emission region has to be considerably magnetized. There the shock acceleration cannot be efficient, or even the shock can hardly happen(e.g., \citealt{2000ApJ...542..235K, 2001MNRAS.328..393A}). We believe that both shock acceleration and electric field acceleration are unlikely to coexist. We do not propose an explanation for why they are happening in the blazar region. However, it is interesting to speculate that these might be a result of the electric field acceleration concentrating in the comparatively smaller volume surrounding the shock and the shock acceleration occurring on around the shock front. In this scenario, the electric field acceleration is best interpreted as a pre-acceleration process in which a fraction of particles obtained energy from the electric field easily to trigger a \emph{Fermi}-type acceleration. Such pre-acceleration process in our steady-state particle spectrum could be simply considered by postulating higher energy for the injected electrons with a Lorentz factor, $\gamma_{0}$.

There are alternative criteria for the blazar classification including the use of the peak frequency of the synchrotron SED component: low synchrotron-peaked blazars (LSP; $\log\nu_{\rm syn, p}<14~\rm Hz$); intermediate synchrotron-peaked blazars (ISP; $14~\rm Hz\leq\log\nu_{\rm syn, p}\leq15~\rm Hz$); and high-synchrotron-peaked blazars (HSP; $\log\nu_{\rm syn, p}>15~\rm Hz$) (e.g., \citealt{1995ApJ...444..567P, 2006A&A...445..441N, 2008A&A...488..867N, 2010ApJ...716...30A, 2016ApJS..226...20F}). It is surprising to find evidence for the migration in the peak positions of the synchrotron SED component during in the activity epochs of Mrk 421. Since most blazars appear as luminous sources characterized by noticeable and rapid flux variability at all observed frequencies, on the basis of the proposed model results, we postulate that the peak frequency of the synchrotron SED component may denote a temporary characteristic of blazars, rather than a permanent one. Taking the finiteness of the sample into account, we defer to this possibility for other blazar sources that are extensively observed.

A potential drawback of the proposed model is that we apply a steady particle distribution instead of time-dependent evolution to model the SEDs of each day and estimate physical parameters in the dissipation region. As a result, it provides an estimation of the temporal evolution of these physical parameters, but we cannot simulate the light curves during the activity epochs. Consequentially, the expected orphan flaring at the X-ray and GeV-TeV $\gamma$-ray bands cannot be generated by the proposed model in the context. On the other hand, in spite of that we introduce a ambient soft photon field to establish the particle distributions, the theoretical photon spectra reproducing in the SSC process results that the proposed model can not to explain the pure orphan $\gamma$-ray flaring. In order to interpret the flares, \cite{2015ApJ...804..111M} proposed the \emph{Ring of Fire} model, where the electrons in the blob up-scatter the ring photons to simulate the origin of orphan $\gamma$-ray flares from the jet dissipation region. Alternatively, the hadronic model has been proposed that either the proton-synchrotron emission or the photohadronic process can explain orphan high energy flares \citep{2015MNRAS.448..910C, 2019ApJ...884L..17S}. In the current context, while the special variability timescale does not be taken into account, the snapshot approach with the steady SSC model allows us to investigate for the temporary evolution of basic physical parameters averaged over a day; this can consider the blobs independently of the difficulties associated with a time dependent model. As time dependent variations of electric and magnetic fields (e.g., \citealt{2018ApJ...853...16K, 2019ApJ...872...65K}) can result in the evolution of the emission electrons, we leave this issue for our future work.

\normalem
\begin{acknowledgements}

We thank the anonymous referee for valuable comments and suggestions.
This work was partially supported by the National Natural Science Foundation of China (Grant Nos. 11673060, 11763005 and 11873043), the Specialized Research Fund for Shandong Provincial Key Laboratory (Grant No. KLWH201804), and the Research Foundation for Scientific Elitists of the Department of Education of Guizhou Province (Grant No. QJHKYZ[2018]068).

\end{acknowledgements}

\bibliographystyle{raa}
\bibliography{msRAA-2020-0138}

\label{lastpage}

\end{document}